\documentclass[journal,oneside,twocolumn,final,letterpaper]{IEEEtran}%
\usepackage{amsfonts}
\usepackage{amsmath}
\usepackage{cite}%
\usepackage{amssymb}%
\usepackage{graphicx}
\begin{document}

\title{Fast Resonance Frequency Modulation in Superconducting Stripline Resonator}
\author{Eran Segev, Baleegh Abdo, \textit{Student Member},\textit{\ IEEE}, Oleg
Shtempluck, and~Eyal Buks~\thanks{This work was supported by MAFAT, Israel
Science Foundation under grant 1380021, the Deborah Foundation and Poznanski
Foundation.}\thanks{The authors are with the Department of Electrical
Engineering and Microelectronics Research Center, Technion, Haifa 32000,
Israel (E-mail: segeve@tx.technion.ac.il).}}
\maketitle

\begin{abstract}
Fast resonance frequency modulation of a superconducting stripline resonator
is investigated. The experiments are performed using a novel device which
integrates a hot electron detector (HED) into a superconducting stripline ring
resonator. Frequency modulation is demonstrated by both applying dc current or
voltage to the HED, and by applying optical illumination, with modulation
frequencies of up to 4.2GHz. Potential applications for such a device are in
telecommunication, quantum cryptography and biofluorescence.

\end{abstract}



\IEEEpeerreviewmaketitle

\section{Introduction}

\PARstart{R}{esonance} parametric amplifiers are characterized by very low
noise, high gain, and phase sensitive amplification. Parametric resonance in
superconducting resonators
\cite{supRes_ParametricResonanceSuperconductingMicro} may allow some
intriguing applications such as quantum
squeezing\cite{Squeezing_PerformanceCavity-parametricAmplif}, quantum
non-demolition measurements \cite{NDM_ComplementarityQuantumNondemolition},
photon creation by the so-called dynamical Casimir
effect\cite{Casimir_GenerationDetectionPhotons}, and more.

Parametric excitation occurs when the resonance frequency of an oscillator
varies in time. The first parametric resonance occurs when the excitation is
performed periodically at twice the resonance frequency $f_{0}$, namely
$f(t)=f_{0}\left[  1+\xi\cos\left(  4\pi f_{0}t\right)  \right]  $
\cite{Mechanics}. The system's response to such an excitation depends on the
dimensionless parameter $\xi Q$ , where $Q$ is the quality factor of the
resonator. When $\xi Q<1$ the system is said to be in the subthreshold regime,
while above threshold, when $\xi Q>1,$ the system breaks into oscillation.
Achieving parametric gain where $\xi Q>1$ requires that the shift in the
resonance frequency exceeds the width of its peak
\cite{supRes_NonlinearMicrowavePropertiesSupercon}.

The frequency modulation mechanism, we employ here, is based on changing the
boundary conditions of a superconducting resonator. This is done by switching
a small section of the resonator to a normal state by using optical
illumination. The switching time in superconductors is usually limited by the
relaxation process of high-energy quasi-particles, also called
'hot-electrons', giving their energy to the lattice, and recombining to form
Cooper pairs. Recent experiments with photodetectors, based on a thin layer of
superconducting Niobium-Nitride (NbN), have demonstrated an intrinsic
switching time on the order of $30%
\mathrm{ps}%
$ and a counting rate exceeding $2%
\mathrm{GHz}%
$ (see \cite{HED_UltrafastSuperconductingSingle-photonDetec} and references
therein). Resonance frequency shift by optical radiation
\cite{supRes_ModulationResonanceFrequency},
\cite{supRes_ObservationBolometricOpticalResponse},
\cite{supRes_FrequencyModulationSuperconductin}, or high-energy particles
\cite{supRes_BroadbandSuperconductingDetectorSuitabl},
\cite{HED_SuperconductingNbNmicrostripDetectors} (for which the required
condition $\xi Q\cong1$ has been achieved) was demonstrated, though no
periodic modulation was reported. Resonance frequency tuning
\cite{NormRR_OnTheStudyOfMicrostripRing} and switching
\cite{NormRR_TheoreticalExperimentalInvestigation} as well as optical and
microwave signal mixing \cite{NormRR_DegenerateParametricAmplification},
\cite{NormRR_ExperimentalInvestigationMicrowave-Optoele} were demonstrated in
normal-conducting GaAs microstrip ring resonators.

In this paper we show experimentally, that resonance frequency modulation, at
twice the resonance frequency, is within reach using superconducting microwave
resonators. Furthermore, the parametric gain threshold conditions, namely $\xi
Q>1,$ is demonstrated in a continuous wave (CW) measurement. The experiments
are performed using a novel device, that integrates a HED into a
superconducting ring resonator. The HED is used as an optically tuned, lumped
element, that changes the boundary conditions of the resonator
\cite{supRes_FrequencyTime-VaryingScatteringPar}, and thus manipulates its
resonance frequencies.

In the following section we describe the circuit design and fabrication
process. The results section starts with the HED response to applied dc
voltage and current, and the resulting effect on the resonance frequencies.
These results are followed by a comparison with a theoretical model.
Afterwards, the effect of CW and modulated infrared (IR)\ light on the
resonance frequency is described, and fast optical modulation of the resonance
frequency is demonstrated.\textbf{\ }

\section{Circuit Design and Fabrication}

\subsection{Circuit Design}

\begin{figure}
[ptb]
\begin{center}
\includegraphics[
height=3.3918in,
width=3.371in
]%
{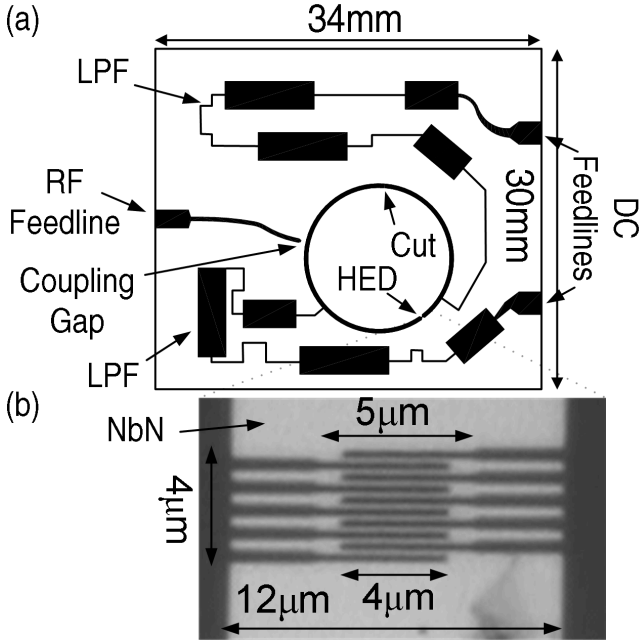}%
\caption{ $\left(  a\right)  $ Device layout. $\left(  b\right)  $ Optical
microscope image of the HED.}%
\label{ResonatorLayoutAndHEDpicture}%
\end{center}
\end{figure}

The circuit layout is illustrated in Fig. \ref{ResonatorLayoutAndHEDpicture}%
(a). The device is made of $8%
\mathrm{nm}%
$ thick NbN stripline, fabricated on a sapphire wafer, with dimensions of
$34\times30\times1%
\mathrm{mm}%
^{3}.$ The design integrates three components. The first is a superconducting
ring resonator and its feedline. Ring configuration is a symmetric and compact
geometry, which is generally suitable for applications, which require
resonance tuning \cite{NormRR_OnTheStudyOfMicrostripRing}. The first few
resonance frequencies are designed for the S\&C bands $\left(  2-8%
\mathrm{GHz}%
\right)  $. The resonator is weakly coupled to its feedline, where the
coupling gap is $0.4%
\mathrm{mm}%
$. The stripline width is set to $347%
\mathrm{\mu m}%
$, to obtain a characteristic impedance of $Z_{0}=50%
\mathrm{\Omega }%
.$

The second component is a HED, which is monolithically integrated into the
ring structure. Its angular location, relative to the feedline coupling
location, maximizes the RF current amplitude flowing through it, and thus
maximizes its coupling to the resonator. The HED, shown in Fig.
\ref{ResonatorLayoutAndHEDpicture}(b), has a $4\times4%
\mathrm{\mu m}%
^{2}$ meander structure, consists of nine NbN superconducting strips. Each
strip has a characteristic area of $0.15\times4%
\mathrm{\mu m}%
^{2}$ and the strips are separated one from another by approximately $0.25%
\mathrm{\mu m}%
$ \cite{HED_Response_TimeNbN}.

The HED operating point is maintained by applying dc bias. The dc\ bias lines,
forming the third component, are designed as two superconducting on-chip
low-pass filters (LPF) with a cut-off frequency of $1.2%
\mathrm{GHz}%
$. As this frequency is lower than the fundamental resonance frequency of the
resonator, the intrinsic fields of the resonator are not appreciably
perturbed. A cut of $20%
\mathrm{\mu m}%
$ is made in the perimeter of the resonator, to force the dc bias current flow
through the HED.

The device, which is top covered by a bare sapphire substrate, is housed in a
gold plated Faraday package made of Oxygen Free High Conductivity (OFHC)
Copper. Superconducting Niobium ground planes are dc-magnetron sputtered on
the inner covers of the package. RF\ power is fed using a SMA launcher,
coupled to the feedline. A dc bias is fed through two $\pi$-LPFs, screwed to
the package, having a cut-off frequency of $1%
\mathrm{MHz}%
$. IR laser light is guided to the device by a fiber optic cable. A through
hole of $1%
\mathrm{mm}%
$ in diameter, is drilled in the Faraday package, and a fiber optic connector
affixes the tip of the fiber cable at approximately $9.55%
\mathrm{mm}%
$ above the HED.%

\subsection{Fabrication Process}

$\ $The fabrication process starts with a thorough pre-cleaning of the
sapphire wafer in solvents. We have experienced that the commonly employed
process of piranha followed by RCA cleaning substantially reduces the NbN
adhesion to the Sapphire wafer. In the next step, $200%
\mathrm{nm}%
$\ thick gold pads are thermally evaporated through a mechanical mask to form
the dc contact pads. The mask partially shadows the evaporation and thus the
pads' perimeters are smoothed. Epitaxy, $8%
\mathrm{nm}%
$ thick, NbN film is then deposited at $700%
\mathrm{{}^{\circ}{\rm C}}%
$ using a dc-magnetron sputtering system
\cite{Fabr_FabricationNanostructuredSuperconducti}. Sputtering parameters are
summarized in table \ref{Sputtering Param} and the process itself is further
detailed in \cite{Baleegh_bifurcation}. Next, an AlN layer of $7%
\mathrm{nm}%
$ thickness is \textit{in-situ} sputtered in N$_{2}$ atmosphere at a
temperature smaller than $100%
\mathrm{{}^{\circ}{\rm C}}%
$. This layer protects the vulnerable NbN layer during the following
fabrication processes and restrain degradation
\cite{Fabr_FabricationCharacterizationUltra-Thin}. It has also a functional
role, as at cryogenic temperatures, it serves as a thermal conducting layer,
which enhances the cooling of the NbN layer. In the next step, the HED meander
stripline is patterned using electron beam lithography (EBL). The deposition
of a $80%
\mathrm{nm}%
$ thick PMMA 950K layer is followed by EBL with the following parameters:
$40$kV$,15$pA, and $2.1$nC/cm, corresponding to acceleration voltage, emission
current, and line dose respectively. Afterwards, the AlN layer is directly
etched through the PMMA mask using ion milling. The remaining AlN layer serves
as a mask for the sequential etching of the NbN layer, using low power
reactive ion etching (RIE) in SF$_{6}$ environment
\cite{Fabr_INVESTIGATION_ETCHING_TECHNIQUES}. The remaining PMMA is removed by
NMP. The last fabrication step is the patterning of the resonator and the LPFs
features. This is achieved by using standard photolithography process. The
photoresist development process (employing AZ-326 photoresist developer), also
wet etches the AlN layer, while the remaining layer is again used as a mask
for the RIE etching of the NbN film.
\begin{table}
\renewcommand{\arraystretch}{1.5}
\caption{Sputtering Parameters}
\centering\begin{tabular} {c | c | c}
\hline\bfseries Process parameter & \bfseries NbN & \bfseries AlN \\
\hline\hline\bfseries Partial flow ratios (Ar,N$_{2}%
$) & (87.5\%,12.5\%)  & (0\%,100\%) \\
\hline\bfseries Base temperature  & $700\,^{\circ}\mathrm{C}$  & $60\,^{\circ
}\mathrm{C}$ \\
\hline\bfseries Base pressure & $ 3.6 \cdot10^{-7} $  torr & $ 1.8 \cdot
10^{-7} $ torr \\
\hline\bfseries Work pressure & $ 6.8 \cdot10^{-3} $  torr & $ 2.9 \cdot
10^{-3} $  torr  \\
\hline\bfseries Discharge current & $360 \mathrm{mA} $  & $360 \mathrm{mA}
$  \\
\hline\bfseries Discharge voltage & $311 \mathrm{V} $  & $292 \mathrm{V}
$  \\
\hline\bfseries Deposition rate & $3.6 \mathrm{\AA}/${sec}  & $1.1 \mathrm
{\AA}/${sec}  \\
\hline\bfseries Thickness ($t$) & $8 \mathrm{nm} $  & $7 \mathrm{nm}$  \\
\hline\bfseries Target-substrate distance & $95 \mathrm{mm} $  & $230 \mathrm
{mm} $ \\
\hline\end{tabular}
\label{Sputtering Param}
\end{table}%

\section{Experimental and Numerical Results}

\begin{figure}
[ptb]
\begin{center}
\includegraphics[
height=1.2531in,
width=3.4835in
]%
{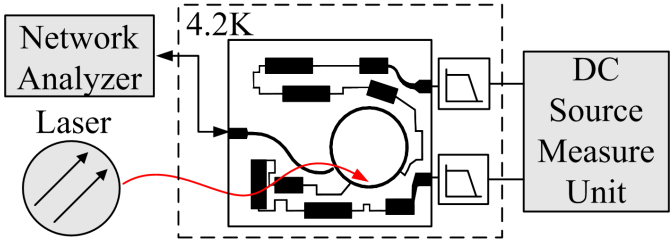}%
\caption{Setup used for reflection measurements.}%
\label{SimpExpSet}%
\end{center}
\end{figure}

All measurements presented in this paper are carried out in a fully immersed
sample in liquid helium ($4.2%
\mathrm{K}%
$). The experimental setup, used for reflection measurements, is schematically
depicted in Fig. \ref{SimpExpSet}. The samples's RF feedline is connected to a
vector network analyzer using a semi-rigid coax cable. The dc feedlines are
connected to a dc source-measure unit using 4-probe wiring. The laser source
has a wavelength of $1550%
\mathrm{nm}%
$.

\subsection{DC I-V Measurements}

\begin{figure}
\centering
\begin{tabular}
[c]{c}%
{\parbox[b]{3.3728in}{\begin{center}
\includegraphics[
height=2.7155in,
width=3.3728in
]%
{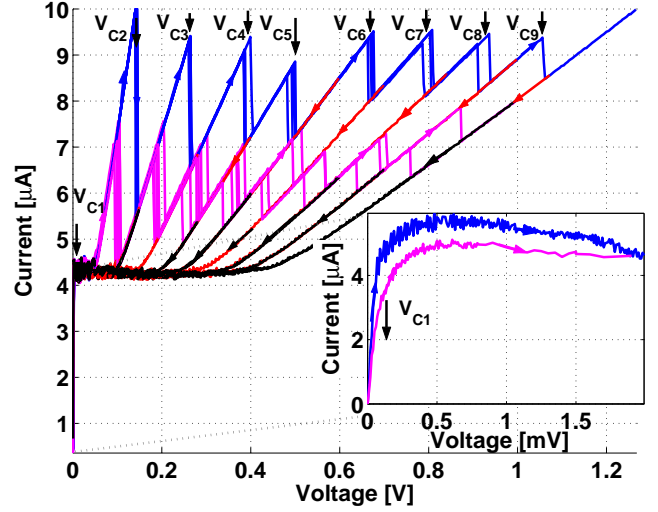}%
\\
(a)
\end{center}}}%
\\%
{\parbox[b]{3.3728in}{\begin{center}
\includegraphics[
height=2.7345in,
width=3.3728in
]%
{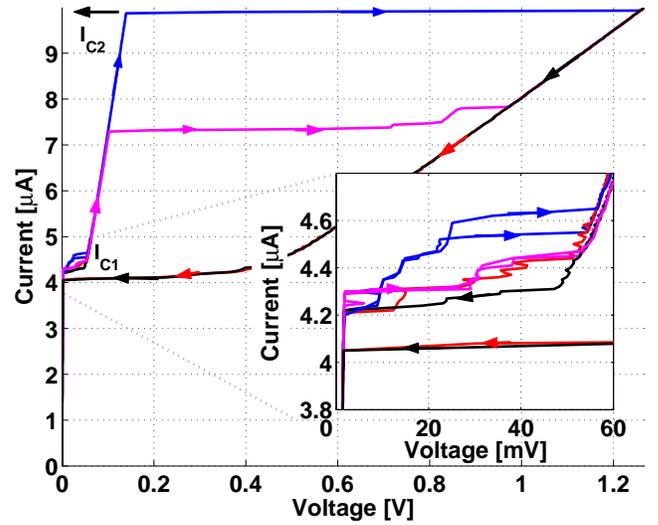}%
\\
(b)
\end{center}}}%
\end{tabular}%
\caption
{Basic I-V characteristic of the HED meander stripline. (a) Current vs. voltage and (b) voltage vs. current measurements. The magenta (blue) curves show an increasing applied voltage and current measurement, with (without) laser illumination. The black (red) curves show a decreasing applied voltage and current measurement, with (without) laser illumination. The insets of (a) and (b) magnify the results for small applied voltages and currents respectively.}%
\label{IVcurves}%
\end{figure}

The basic I-V characteristics of the HED\ meander stripline, shown in Fig
\ref{IVcurves}, exhibit a highly complex hysteretic behavior. Panel
\ref{IVcurves}(a) plots nine current measurements, obtained while increasing
the applied voltage (blue), one on top of the other, where each measurement
starts at zero voltage and ends at a different maximum voltage, slightly above
the voltages $V_{Cn},$ depicted in the figure. The corresponding nine current
measurements, obtained while decreasing the applied voltage are also plotted
(red). The magenta and black curves correspond to similar measurements, taken
while the HED is being illuminated. The measurements for low applied voltages
and currents are enlarged in the insets of Fig \ref{IVcurves}(a) and (b)
respectively, where the finite resistance is due to the contacts.

At panel \ref{IVcurves}(a), nine, clearly distinguished, abrupt jumps in the
measured current are noticed. The number of current jumps corresponds to the
number of stripline sections that compose the meander shape of the HED. Each
jump is the result of a large increase in the HED resistance due to a
transition of one section from the superconducting state to the normal one.
This behavior is typical for a superconducting microbridge and is caused by
the formation of a hotspot in the bridge area
\cite{Self-heatingNormalMetalsSuperconductors}. Each critical voltage
$V_{Cn},$ varies, in general, between different scans, indicating thus, the
stochastic nature of the transitions between bistable states. The fluctuation
$\Delta V_{Cn}$ in $V_{Cn},$ between different scans, characterizes the
lifetime of the pre-jump metastable states of the system. The increase in
$\Delta V_{Cn}$ at high voltages indicates a decrease in the lifetime of
metastable states because of larger temperature fluctuations. The combined
results of the increasing and decreasing applied voltage measurements show
that large hysteresis is present at all current jumps except for the first
one. This observation indicates, that only one section at a time can be biased
into subcritical conditions. Furthermore, only the section responsible for the
first jump, at $V_{C1}$, doesn't suffer form hysteresis and thus can
repetitively respond to radiation. Probably, the cause for this discretization
of the critical current is the non-uniformity in the meander shape of the HED
\cite{Self-heatingNormalMetalsSuperconductors},
\cite{HED_TimeDelayResistive-stateFormation}. Our finding clearly shows that
the non-uniformity may substantially reduces the effective area of the HED, up
to a fraction of $1/9$ of its printed area.

The same measurements are repeated while constantly illuminating the HED with
approximately $27%
\mathrm{nW}%
$ IR laser. In these measurements the current jumps occur at lower applied
voltages, $V_{Cn}^{^{\prime}}<$ $V_{Cn}$. In addition, $\Delta V_{Cn}$ are
substantially widened. Although the decrease in the critical voltage values
can be explained by local heating due to the IR\ illumination, the increase in
$\Delta V_{Cn}$, especially at low voltages, imply that photon absorptions
cause discrete events that considerably increase the instability of the HED.

Panel \ref{IVcurves}(b) shows voltage measurements, obtained while increasing
the applied current, with (magenta) and without (blue) IR\ illumination, and
while decreasing the applied current, with (black) and without (red)
IR\ illumination. Two abrupt voltage jumps occur at distinguishable critical
currents of $I_{C1}\cong$ $4.2%
\mathrm{\mu A}%
$ and $I_{C2}\cong9.8%
\mathrm{\mu A}%
$. IR illumination has a negligible measured effect on $I_{C1}$ and a strong
effect on $I_{C2}$ values. All voltage jumps suffer from hysteresis and
therefore current bias is an unsuitable method for repetitive radiation
detection.

\subsection{DC I-V Effect on the Resonance Lineshape}

\begin{figure}
[ptb]
\begin{center}
\includegraphics[
height=2.4111in,
width=3.3728in
]%
{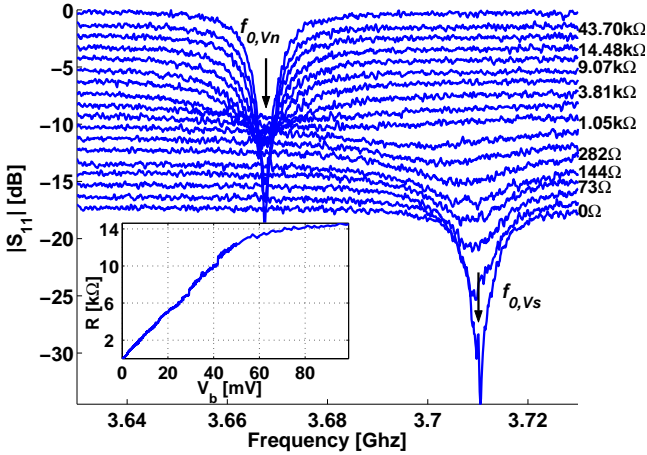}%
\caption{Several $|S_{11}|$ measurements of the second resonance mode, as a
function of frequency, for various HED resistance values. The measurements are
obtained while applying variable voltage and the HED resistance is measured
using 4-probe technique. Plots are shifted vertically for clarity. The inset
shows the resistance vs. voltage characteristics of the HED. }%
\label{S11VsRes}%
\end{center}
\end{figure}

Fig. \ref{S11VsRes} shows several $|S_{11}|$ measurements as a function of
frequency, in the vicinity of the second resonance mode, for various HED
resistance values. For clarity, the resonance curves are vertically shifted
upwards, for increasing resistance values. The measurements are obtained while
applying variable voltage $V_{b}$, and the resistance is measured
simultaneously with the $|S_{11}|$ data using standard 4-probe technique. The
RF input power is set to $-64.7$dBm, where the resonator is in the linear
regime \cite{baleegh_NonlinearDynamicsResonanceLines}. The inset of Fig.
\ref{S11VsRes} plots the measured HED resistance as a function of $V_{b} $.

The dependence of the resonance characteristics on the HED resistance
$R_{\text{HED}},$ can be described as followed. At zero applied voltage the
resonance frequency is $f_{0,Vs}\cong3.71%
\mathrm{GHz}%
.$ At very low voltages, as the HED is biased far below critical conditions,
its resistance is negligible and its influence on the resonance curve as well.
As the resistance increases, the resonance frequency slightly red shifts, and
more important, the \textit{Q}-factor is significantly reduced due to
dissipation in the HED. This behavior continues up to a point, at
$R_{\text{HED}}\cong1k\Omega,$ where the resonance curve can be hardly
detected. When increasing the resistance beyond that point the trend of the
\textit{Q}-factor changes, the dissipation decreases, and the resonance curve
reemerges at a new resonance frequency, $f_{0,Vn}\cong3.665%
\mathrm{GHz}%
$, red shifted by approximately $45%
\mathrm{MHz}%
\ $relative to its original value. The new resonance \textit{Q}-factor has a
value similar to the original one. When further increasing the resistance, the
trend of the \textit{Q}-factor continues but no additional resonance shift
occurs. The behavior of the \textit{Q}-factor suggests that as $R_{\text{HED}%
}$ increases, the RF current amplitude of the resonance mode in the HED is
reduced, due to current redistribution, and thus the total power dissipation decreases.

\begin{figure}
[ptb]
\begin{center}
\includegraphics[
height=2.7302in,
width=3.3728in
]%
{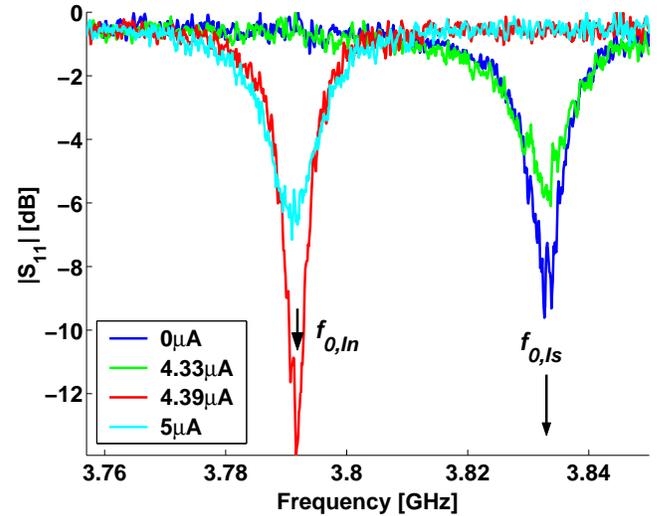}%
\caption{Several $|S_{11}|$ measurements of the second resonance mode, as a
function of frequency. The measurements are obtained while applying variable
current.}%
\label{S11VsCurrent}%
\end{center}
\end{figure}

Similar behavior, with one major exception, can be observed under applied
current, as shown in Fig \ref{S11VsCurrent}. The blue, green, red, and cyan
curves are taken with subcritical $0%
\mathrm{\mu A}%
$, $4.33%
\mathrm{\mu A}%
<I_{C1}$, and over critical $4.39%
\mathrm{\mu A}%
$, $5%
\mathrm{\mu A}%
>I_{C1}$ applied currents, respectively, where $I_{C1}$ is the current at
which a first jump in the measured voltage occurs. There are two well defined
resonance frequencies, $f_{0,Is}=3.83%
\mathrm{GHz}%
$, and $f_{0,In}=3.79%
\mathrm{GHz}%
,$ which corresponds to applied currents below, and above $I_{C1}$
respectively. $f_{0,Is}$, and $f_{0,In}$ slightly differ from $f_{0,Vs}$, and
$f_{0,Vn}$ as the two measurements were taken at different thermal cooldown
cycles. Low \textit{Q}-factor curves are absent from this measurement because
under applied current, the HED can not be biased into intermediate resistance
values. At bias currents below $I_{C1},$ the HED has low resistance, which
only slightly increases as the current approaches $I_{C1}$. As a result, no
resonance shift occurs, and only the \textit{Q}-factor slightly reduces as the
current increases. This behavior changes abruptly once the HED resistance
crosses a rather low, critical value, $R_{C} $. A self-sustained hotspot is
generated \cite{Self-heatingNormalMetalsSuperconductors}, quickly expends, and
the HED becomes resistive. $I_{C1}$ is the bias current at which
$R_{\text{HED}}=R_{C}$ is obtained. This thermal runaway causes an abrupt red
shift of $\Delta f_{0}\cong40%
\mathrm{MHz}%
$ in the resonance frequency. Further increase of the bias current beyond
$I_{C1}$ increases the power dissipation and heat generation in the HED. This
increases the local temperature and dissipation near the HED, and thus causes
\textit{Q}-factor reduction.%

\subsection{Resonance Frequency Shift Modeling}

To account for our results we calculate the resonance characteristics of our
device, as a function of HED resistance.

\begin{figure}
[ptb]
\begin{center}
\includegraphics[
height=1.5627in,
width=3.4835in
]%
{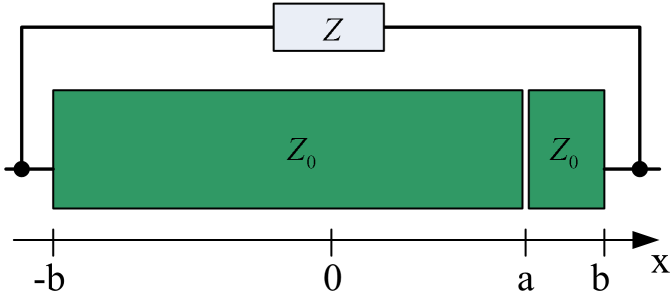}%
\caption{Resonator model.}%
\label{modelblockdiag}%
\end{center}
\end{figure}

As shown in Fig. \ref{modelblockdiag}, the ring resonator is modeled as a
straight transmission line, extending in the $\pm x$ directions. The HED is
represented by a lumped discontinuity, $Z=R+jwL$, connecting $x=\pm b$ points
together, where $R$ is the resistance, $\omega$ is the angular frequency, and
$L$ is the total inductance of the meander shape of the HED. The transmission
line has a cut at point $x=a.$ The couplings to the RF and dc feedlines are
neglected.%

The voltage along the resonator's transmission line is given by a standing
waves expression of the form \cite{Model_PlanarMicrostripRingResonatorFilters}%
\[
V\left(  x\right)  =\left\{
\begin{array}
[c]{lc}%
A\cos\left(  \beta x\right)  +B\sin\left(  \beta x\right)  & -b<x<a\\
C\cos\beta\left(  x-a\right)  +D\sin\beta\left(  x-a\right)  & a<x<b
\end{array}
\right.  ,
\]

where $\beta=2\pi f\sqrt{\epsilon_{r}}/c$ is the propagation constant along
the transmission line, $f$ \ is the frequency, $\epsilon_{r}$ is the relative
dielectric constant, and $c$ is the speed of light in vacuum.

The current is given by $I\left(  x\right)  =\left(  i/\beta Z_{0}\right)
dV/dx, $ where $Z_{0}$ is the characteristic impedance of the line. By
applying the following boundary conditions: $\left(  1\right)  $ $I\left(
a_{-}\right)  =I\left(  a_{+}\right)  =0,$ $\left(  2\right)  $ $I\left(
-b\right)  =I\left(  b\right)  $, and $\left(  3\right)  $ $V\left(  b\right)
-V\left(  -b\right)  =I\left(  b\right)  Z,$ we easily derive a boundary
condition equation which yields the complex resonance frequencies of the resonator.

A phase transition of $Z$ from superconducting to normal state simultaneously
causes changes in its resistive, $\Delta R>0$, and inductive, $\Delta L>0,$
parts. Both changes contribute to a resonance shift in the same direction. For
a very thin superconducting films $\Delta R\gg\Delta L$ resulting in $\Delta
Z\cong\Delta R.$ Three fitting parameters are used in the model. Best fit
results are obtained for $R=0\Omega,$ $Z_{0}=55\Omega,$ and $\omega
L=14.5Z_{0}$, which leads to $L=37.3$nH for the second resonance mode. $R\ll
Z_{0}$ in the superconducting state, and thus negligible. The characteristic
impedance $Z_{0}$ value is in a very good agreement with the designed value of
50$\Omega.$ The calculated inductance $L$ of the meander line at $4.2%
\mathrm{K}%
$ without any applied current is $6.05$nH
\cite{KinInd_CurrentControlledVariableDelay}, but this value is strongly
dependant on temperature and current density, so the fit parameter value is in
reasonable agreement with the calculated one
\cite{KinInd_CurrentControlledVariableDelay}%
,\cite{KinInd_CurrentTemperatureControlledVariabl}.

Table \ref{ResFreqChar} summarizes the first three solutions of the boundary
condition equation for the two extreme cases of $\Delta R=0$ and $\Delta
R\rightarrow\infty$. Comparing these results to the measured results (taken at
two different cooldown cycles), also summarized in table \ref{ResFreqChar},
shows a good agreement, especially for the second and third modes, for which
the resonator is designed.%
\begin{table*}
\renewcommand{\arraystretch}{1.5}
\caption{Resonance Frequency Characteristics}
\centering\begin{tabular} {c || c | c | c ||c | c | c}
\hline& \multicolumn{3}{c||}{\bfseries Numerical Results} &
\multicolumn{3}{c}{\bfseries Experimental Results} \\
\hline${\footnotesize{n}}$ & $\begin{array}{c}
{\footnotesize f}_{0}\text{{\footnotesize\ [GHz] }} \\
{\footnotesize\Delta R=0}\end{array}$ & $\begin{array}{c}
{\footnotesize f}_{0}\text{{\footnotesize\ [GHz] }} \\
{\footnotesize\Delta R\rightarrow\infty}\end{array}$ & ${\footnotesize\Delta
f}_{0}${\footnotesize\ }$\text{{\footnotesize\
[MHz]}}$ & $\begin{array}{c}
{\footnotesize f}_{0}\text{{\footnotesize\ [GHz] }} \\
{\footnotesize\Delta R=0}\end{array}$ & $\begin{array}{c}
{\footnotesize f}_{0}\text{{\footnotesize\ [GHz] }} \\
{\footnotesize\Delta R\rightarrow\infty}\end{array}$ & ${\footnotesize\Delta
f}_{0}\text{{\footnotesize\ [MHz]}}$ \\
\hline\hline\bfseries1& 1.913  & 1.873 & 39.8 & 1.59 & 1.58 & 10 \\
\hline\bfseries
2& 3.791 & 3.747 & 43.7 & 3.874, 3.711 & 3.829, 3.668 & 45, 43 \\
\hline\bfseries3& 5.654 & 5.62 & 34.1 & 5.634, 5.38 & 5.608, 5.35 & 26, 30 \\
\end{tabular}
\label{ResFreqChar}
\end{table*}%

\begin{figure}
[ptb]
\begin{center}
\includegraphics[
height=2.6161in,
width=3.3728in
]%
{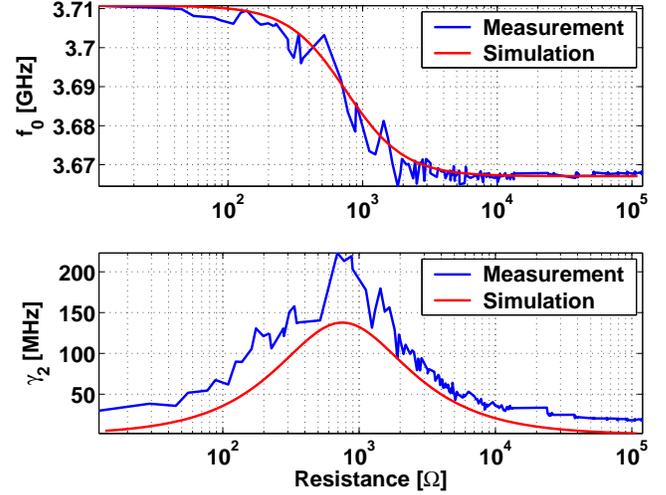}%
\caption{Resonance frequency and unloaded damping rate $\gamma_{2}$ of the
second resonance mode.}%
\label{centfreq_qfactor}%
\end{center}
\end{figure}

Fig. \ref{centfreq_qfactor} shows the second resonance characteristics,
resonance frequency and unloaded damping rate $\gamma_{2}$
\cite{Squeezing_PerformanceCavity-parametricAmplif}, of the experimental data
(blue) and the numerically calculated data (red). The rate $\gamma_{2}$ is
extracted from the data plotted in Fig. \ref{S11VsRes} using the method
presented in the appendix. The upper subplot shows the resonance frequency as
a function of $\Delta R.$ The calculated resonance frequency, at zero
resistance, is $f_{0}=3.791%
\mathrm{GHz}%
$, which equals the mean value of the resonance frequency measured at
different cooldown cycles. In this subplot, the calculated data is corrected
by $-80%
\mathrm{MHz}%
$ to overlap between the first calculated and measured point. Both curves show
the same dependence on $\Delta R.$ The lower subplot shows the unloaded
damping rate $\gamma_{2},$ as a function of $\Delta R$. Also in this case, a
good agreement with the experiment is obtained, and as expected, the measured
damping rate exceeds the calculated one, due to losses, which are not taken
into account in the model.%

\begin{figure}
\centering
\begin{tabular}
[c]{cc}%
{\parbox[b]{1.6561in}{\begin{center}
\includegraphics[
height=1.3145in,
width=1.6561in
]%
{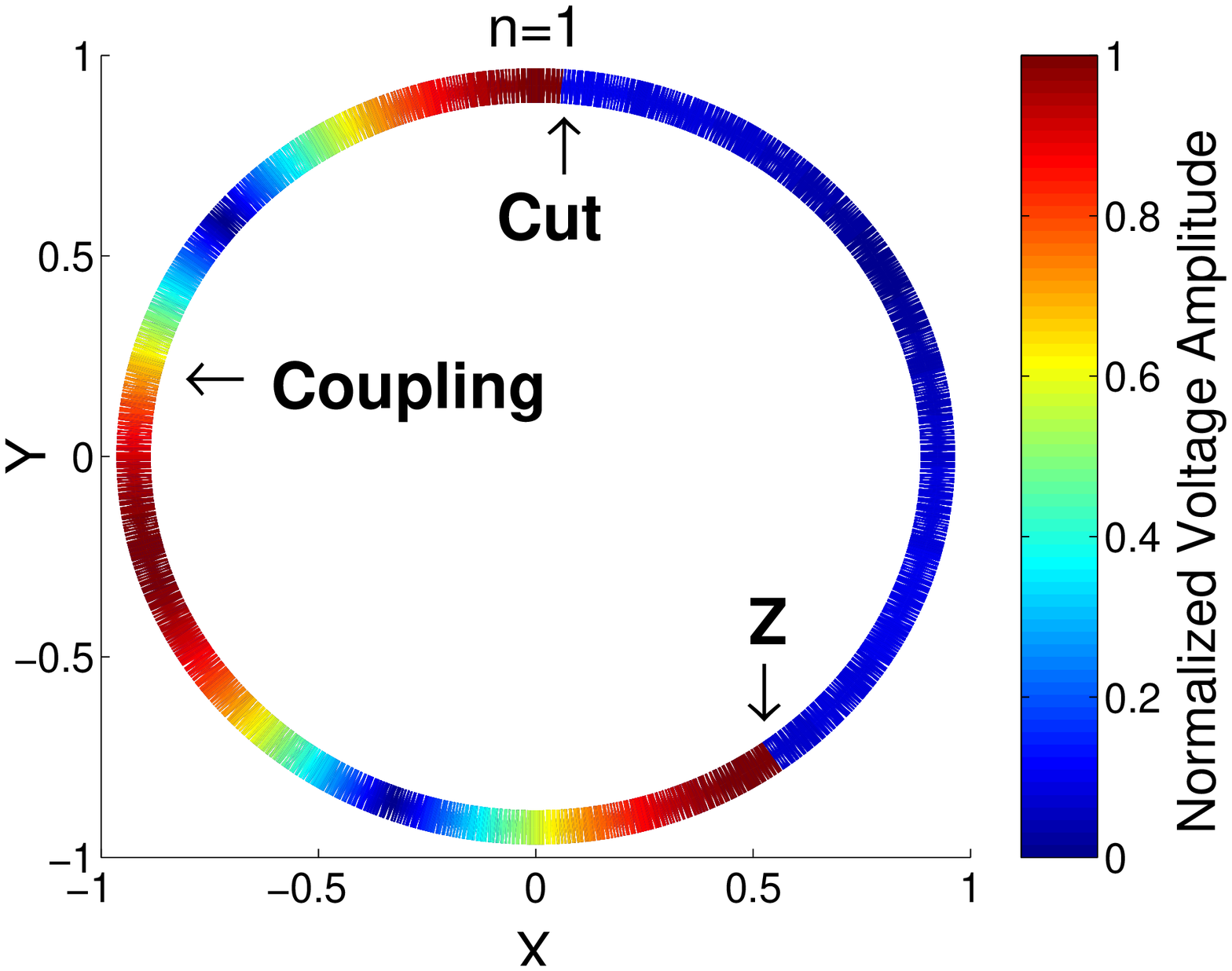}%
\\
(a)
\end{center}}}%
&
{\parbox[b]{1.6561in}{\begin{center}
\includegraphics[
height=1.3145in,
width=1.6561in
]%
{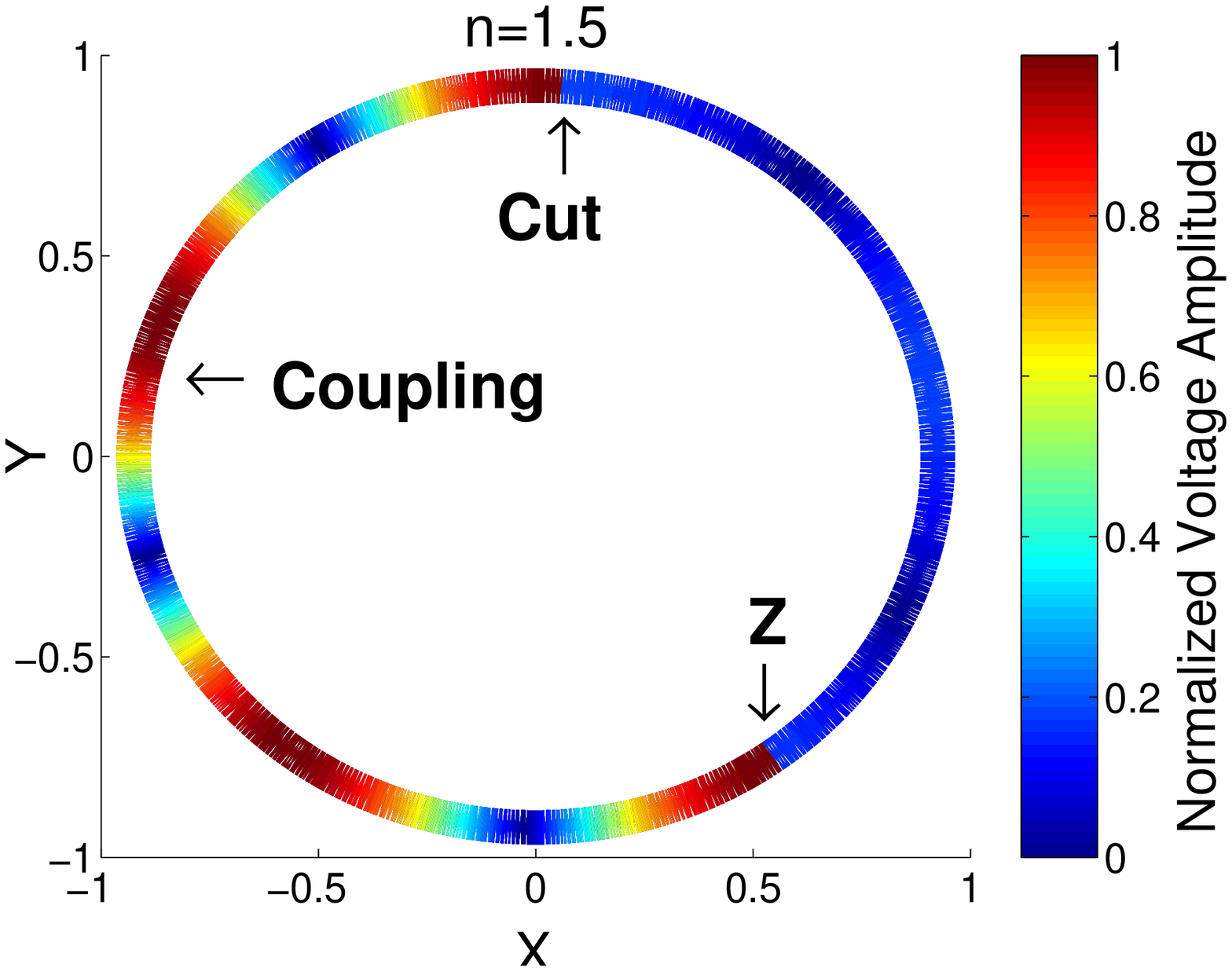}%
\\
(b)
\end{center}}}%
\end{tabular}%
\caption
{Normalized voltage amplitudes as a function of the ring's angular location, for the (a) second and (b) third resonance modes.}%
\label{Ring Volt Dist}%
\end{figure}

The coupling between the different modes and the HED can be characterized by
the current amplitude through the HED. The model predicts normalized current
amplitudes of $7.3\%$ and $5.7\%$ inside the lumped element, for the second
and third modes respectively. This rather weak coupling is the result of the
rather high kinetic inductance of the HED. To estimate the coupling of modes
two and three to the feedline \cite{FoundationsForMicrowaveEngineering}, we
show in Fig. \ref{Ring Volt Dist} the normalized voltage amplitudes, as a
function of the ring's angular location. The calculated normalized voltage
amplitudes, at the feedline coupling location, are $71\%$ and $92\%,$
respectively. The voltage amplitudes distribution have, in general, a strong
dependence on the resonance frequency, and hence on $\Delta R$, but because of
the rather small resonance shift, the voltage amplitudes at the coupling
location change by less than $2\%$.

\subsection{IR Illumination Effect on the Resonance Lineshape}

\begin{figure}
[ptb]
\begin{center}
\includegraphics[
height=2.7164in,
width=3.3728in
]%
{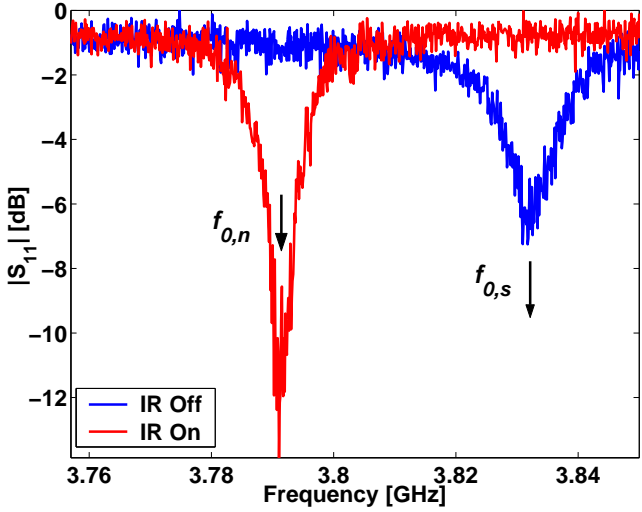}%
\caption{Several $|S_{11}|$ measurements of the second resonance mode, as a
function of frequency, while applying sub-critical current, with (red) and
without (blue) IR illumination.}%
\label{S11VsLaser}%
\end{center}
\end{figure}

Fig. \ref{S11VsLaser} plots $|S_{11}|$ \ measurements with (red) and without
(blue) IR illumination. The effective IR\ illumination power, impinging on the
HED, is $\ $approximately $27%
\mathrm{nW}%
.$ The RF input power is set to $-64.7$dBm and the HED is biased with a
subcritical dc current of $4.14%
\mathrm{\mu A}%
,$ which only weakly influences the resonance curve. When the illumination is
turned on, the resonance frequency abruptly shifts to a lower frequency. The
new resonance lineshape has the same characteristics as the resonance
lineshape measured without illumination under supercritical bias current of
$I=4.39%
\mathrm{\mu A}%
>I_{C1}.$ This measurement clearly shows that the resonance frequency is
sensitive to IR illumination. The measured results in this experiment yield
$\xi Q\cong4.14.$

\subsection{Fast Modulation of the Resonance Frequency}

\begin{figure}
[ptb]
\begin{center}
\includegraphics[
height=1.8542in,
width=3.4532in
]%
{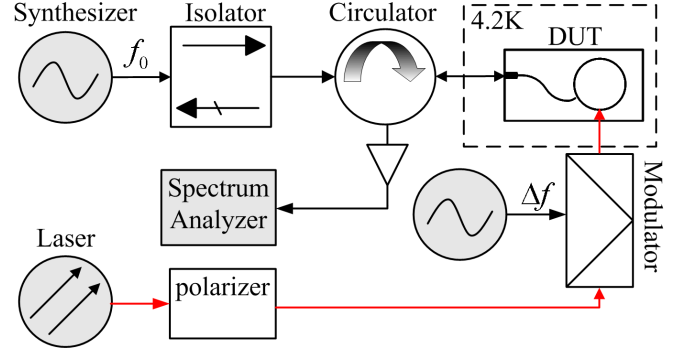}%
\caption{Setup used for frequency modulation.}%
\label{laserexpsetup}%
\end{center}
\end{figure}

Fast modulation of the resonance frequency is performed using the experimental
setup depicted in Fig \ref{laserexpsetup}. The resonator is excited by a CW
pump signal, at frequency $f_{0}=3.71%
\mathrm{GHz}%
,$ which coincides with the second resonance frequency. The optical signal is
modulated at frequency $\Delta f$, using a Mach-Zener modulator driven by a
second CW signal, phase locked with the first one. The reflected power is
mainly composed of three tones. One is the reflected pump signal at frequency
$f_{0}$. The other two are sidebands, produced by mixing the pump signal and
the optical modulation signal, and are found at frequencies $f_{1,2}=f_{0}%
\pm\Delta f.$ Occasionally, higher orders of the mixed signals are also
detected. The amplified reflected power is measured using a spectrum analyzer,
which tracks the $f_{1}=f_{0}+\Delta f$ tone. No dc bias is needed in this
measurement as the RF probe signal also serves as a bias signal for the HED.
This bias scheme has two major advantages over the dc bias scheme; first the
RF pump bias signal has lower noise, as the $1/f$ and line noises are avoided.
Second, the RF bias signal introduces a strong non-linear mechanism
\cite{baleegh_NonlinearDynamicsResonanceLines},
\cite{Baleegh_IntermodulationGainNonlinearNbN}, which will be discussed in a
future publication \cite{segev_ToBePublished}, and produces a high internal
gain of the induced optical signal.%

\begin{figure}
[ptb]
\begin{center}
\includegraphics[
height=2.591in,
width=3.3728in
]%
{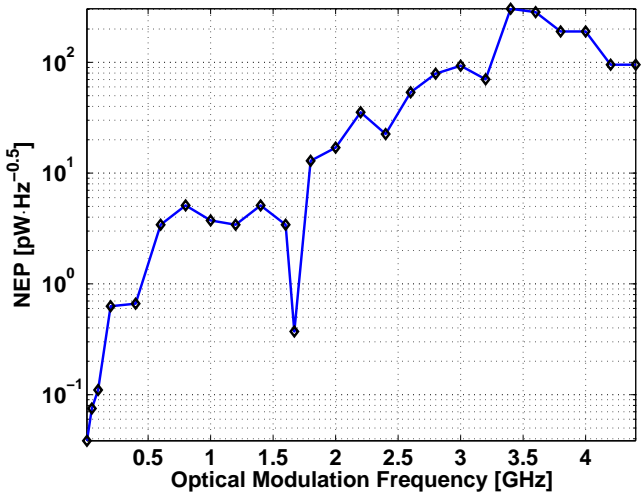}%
\caption{NEP for various optical modulation frequencies.}%
\label{NEP}%
\end{center}
\end{figure}

The device is not designed for radiation detection. Nevertheless, we find it
useful to characterize the response to optical modulation by its noise
equivalent power (NEP). Fig. \ref{NEP} shows the NEP of the device for various
optical modulation frequencies. Each NEP data point is derived out of several
reflected power measurements in the vicinity of $f_{1}$, where each
measurement is performed with a different pump power, ranges between $-54.7$
dBm and $-45.7$ dBm. The best NEP is measured for a modulation frequency of
$\Delta f=10%
\mathrm{MHz}%
$ and equals $38$fW/$\sqrt{%
\mathrm{Hz}%
}$. A rather low NEP of $370$fW$/\sqrt{%
\mathrm{Hz}%
}$ is measured for a modulation frequency of $\Delta f=1.699%
\mathrm{GHz}%
.$ For this modulation frequency, $f_{1}$ coincides with the third resonance
frequency of the resonator. Higher modulation frequencies than the ones
presented in Fig. \ref{NEP} are also observed, but the measured NEP is
relatively poor, mainly due to the bandwidth limitations of the microwave
components composing the experimental setup.%

\section{ Discussion}

Fast modulation of the resonance frequency is experimentally demonstrated.
Furthermore, the parametric gain threshold condition is achieved in a CW
measurement. The main problem, that currently prevents parametric gain to
occur, is the relatively low photon flux that impinges the HED. Due to losses
along the optical path, especially the expansion of the Gaussian beam from the
tip of the fiber to the HED, the largest photon flux we currently manage to
apply is approximately $13$ photons per modulation cycle, at twice the
resonance frequency. Taking into account the effective area of the HED and its
quantum efficiency \cite{HED_SensitivityGigahertzCountingPerformanc} we
estimate that the optical power flux is two orders of magnitude lower than the
threshold power. Therefore parametric gain can not be achieved using the
present device. Future devices will address this problem.

\section{Conclusion}

We have reviewed the design, fabrication, and measurement results of our
optically tunable resonator. Resonance frequency modulation frequency as high
as $4.2%
\mathrm{GHz}%
$ is achieved, and a NEP of 38fW/$\sqrt{%
\mathrm{Hz}%
}$ in the IR is measured. The parametric gain threshold condition is achieved
in a CW measurement. Moreover, the results are shown to be in a good agreement
with a theoretical modeling. The approach used in this device has a great
potential of achieving a parametric amplification in superconducting resonators.%

\appendix[Damping Rare Extraction]%
The universal expression for the reflection amplitude of a linear resonator
near resonance is \cite{Squeezing_PerformanceCavity-parametricAmplif},
\cite{Baleegh_Nonlinear_Coupling}%

\begin{equation}
S_{11}=\frac{i\Omega+\left(  \gamma_{1}-\gamma_{2}\right)  }{i\Omega-\left(
\gamma_{1}+\gamma_{2}\right)  },\label{S11}%
\end{equation}
where $\Omega=\omega_{p}-\omega_{0}$ is the pump angular frequency $\omega
_{p}$, relative to the angular resonance frequency $\omega_{0}$, $\gamma_{1}$
is the coupling constant between the resonator and the feedline, and
$\gamma_{2}$ is the unloaded damping rate of the resonance. The damping rate
is numerically extracted by expanding Eq. (\ref{S11}) to first order in
$\Omega$,%

\[
S_{11}=r_{0}+r_{1}\Omega+O\left(  \Omega^{2}\right)  ,
\]
where $r_{0}$ is the $S_{11}$ value at the resonance frequency,%

\[
r_{0}=\frac{\gamma_{2}-\gamma_{1}}{\gamma_{1}+\gamma_{2}},
\]
and $r_{1}$ is the slope of the imaginary part of $S_{11},$%

\[
r_{1}=-i\frac{2\gamma_{1}}{\left(  \gamma_{1}+\gamma_{2}\right)  ^{2}}.
\]
Note that the extraction of $r_{1}$ is less accurate for low \textit{Q}-factor
curves, and thus the calculated loss factor suffers from a rather large
impreciseness at that regime.


\bibliographystyle{IEEEtran}
\bibliography{IEEEabrv,eran}
%

\begin{biographynophoto}{Eran Segev-Arbel}
was born in Haifa, Israel in 1975. He received the B.Sc. degree in electrical
engineering from the Technion--Israel Institute of Technology, Haifa, Israel,
in 2003. He is currently working toward the M.Sc. in electrical engineering at
the Technion. His research is focused on parametric gain in superconducting
microwave resonators.%
\end{biographynophoto}%
%

\begin{biographynophoto}{Baleegh Abdo}
(S'2002) was born in Haifa, Israel in 1979. He received the B.Sc. degree in
computer engineering, in 2002, and the M.Sc. degree in electrical engineering
in 2004, both from the Technion--Israel Institute of Technology, Haifa,
Israel. Currently he is pursuing the Ph.D. degree in electrical engineering at
the Technion. His graduate research interests are nonlinear effects in
superconducting resonators in the microwave regime, resonator coupling and
quantum computation.%
\end{biographynophoto}%
%

\begin{biographynophoto}{Oleg Shtempluck}%
\textbf{\ }was born in Moldova in 1949. He received the M.Sc. degree in
electronic engineering from the physical department of Chernovtsy State
University, Soviet Union, in 1978. His research concerned semiconductors and
dielectrics. From 1983 to 1992, he was a team leader in the division of design
engineering in Electronmash factory, and from 1992 to 1999 he worked as stamp
and mould design engineer in Ikar company, both in Ukraine. Currently he is
working as a laboratory engineer in Microelectronics Research Center,
Technion- Israel Institute of Technology, Haifa, Israel.%
\end{biographynophoto}%
%

\begin{biographynophoto}{Eyal Buks}
received the B.Sc. degree in mathematics and physics from the Tel-Aviv
University, Tel-Aviv, Israel, in 1991 and the M.Sc. and Ph.D. degrees in
physics from the Weizmann Institute of Science, Israel, in 1994 and 1998,
respectively. His graduate work concentrated on interference and dephasing in
mesoscopic systems. From 1998 to 2002, he worked at the California Institute
of Technology (Caltech), Pasadena, as a Postdoctoral Scholar studying
experimentally nanomachining devices. He is currently a Senior Lecturer at the
Technion---Israel Institute of Technology, Haifa. His current research is
focused on nanomachining and mesoscopic physics.%
\end{biographynophoto}%




\bigskip
\end{document}